\documentclass[preprint,showpacs,amsmath,amssymb]{revtex4}
\usepackage{amssymb}
\usepackage{amsmath}
\usepackage{graphicx}
\usepackage{epsfig}
\usepackage{subfigure}
\usepackage{amsfonts}
\usepackage{CJK}

\begin{document}

\title{Transferring quantum entangled states between multiple single-photon-state qubits and coherent-state qubits in circuit QED}
\author{Qi-Ping Su$^{1}$}
\author{Hanyu Zhang$^{1}$}
\author{Chui-Ping Yang$^{1,2}$}
\email{yangcp@hznu.edu.cn}
\address{$^1$Department of Physics, Hangzhou Normal University, Hangzhou, Zhejiang 311121, China}
\address{$^2$Quantum Information Research Center, Shangrao Normal University, Shangrao, Jiangxi 334001, China}

\begin{abstract}
We present a way to transfer maximally- or partially-entangled states of $n$ single-photon-state (SPS) qubits
onto $n$ coherent-state (CS) qubits, by employing $2n$ microwave cavities coupled to a
superconducting flux qutrit. The two logic states of a SPS qubit here are represented by the
vacuum state and the single-photon state of a cavity, while the two logic states of a CS qubit are encoded with two coherent states of
a cavity. Because of using only one superconducting qutrit as the coupler, the circuit architecture is significantly
simplified. The operation time for the state transfer does not increase with the increasing of the number of qubits. When the dissipation of the system is negligible, the quantum state can be transferred in a deterministic way since no measurement is required. Furthermore, the higher-energy intermediate level of the coupler qutrit is not excited during the entire operation and thus decoherence from the qutrit is greatly suppressed. As a specific example, we numerically demonstrate that the high-fidelity transfer of a Bell state of two SPS qubits onto two CS qubits is achievable within the present-day circuit QED technology. Finally, it is worthy to note that when the dissipation is negligible, entangled states of $n$ CS qubits can be transferred back onto $n$ SPS qubits by performing reverse operations. This proposal is quite general and can be extended to accomplish the same task, by employing a natural or artificial atom to couple $2n$ microwave or optical cavities.
\end{abstract}
\maketitle
\date{\today }

\section{Introduction}

Discrete- and continuous-variable methods to optical quantum communication
and quantum information processing (QIP) rely on different encoding qubits
for their realization [1-3]. For the discrete-variable approaches, single
photons are involved [4,5], the photonic qubits live in a two-dimensional
Hilbert space formed, for instance, by orthogonal polarizations or the
absence and presence of a single photon. For convenience, we define such
photonic qubits as single-photon-state (SPS)\ qubits. In contrast, for the
continuous-variable approaches, encoding is implemented in the quadrature
components of a light field with an inherently infinite-dimensional space.
Quantum information carried by a qubit can be indicated as an arbitrary
superposition of coherent states or an arbitrary superposition of cat states
of light waves [6,7]. In recent years, besides progress achieved with SPS
qubits, considerable theoretical and experimental activity have also been
shifted to QIP with coherent-state (CS)\ or cat-state (C-S)\ qubits. There
exists a number of works for QIP using qubits encoded with coherent states
[6,8-10] or encoded with cat-state qubits [7,11-16].

Both encodings of SPS qubits and CS/C-S qubits have advantages and drawbacks
[17]. CS/C-S qubits benefit from unconditional operations, unambiguous state
discrimination, high detection efficiencies, and more practical interfacing
with conventional information technology. However, it is commonly recognized
that quantum states of multiphoton wave fields are difficult to manipulate
and suffer from intrinsically limited fidelities. On the other hand, quantum
states of SPS qubits are relatively easy to operate, but usually suffer from
decoherence due to the loss of a single photon. A combination of qubits with
different encodings, that is, in a hybrid architecture [18], may offer great
advantages [19]. Some operations may be more efficient using SPS qubits
while others may better take advantage of CS/C-S qubits. In this endeavour,
it is a key requirement to transfer information or entanglement between
qubits with different encodings.

The goal of this work is aimed at transferring the following type of
entangled state
\begin{equation}
c\left\vert i_{1}\right\rangle \left\vert i_{2}\right\rangle ...\left\vert
i_{n}\right\rangle +d\left\vert \overline{i}_{1}\right\rangle \left\vert
\overline{i}_{2}\right\rangle ...\left\vert \overline{i}_{n}\right\rangle
\end{equation}%
between $n$\ SPS qubits and $n$\ CS qubits. For $c\neq d,$ the state (1) is
a \textit{partially-entangled} state. For $c=d=1/\sqrt{2}$, the state (1) is
a \textit{maximally-entangled} state, which is called Bell state when $n=2$\
or Greenberger--Horne--Zeilinger (GHZ) state when $n\geq 3.$ As is well
known, Bell states and GHZ states are crucial resources and play an
important role in quantum communication and QIP. On the other hand,
transferring entangled states between qubits of different encodings is of
fundamental interest in quantum mechanics, and may be necessary and
important in the hybrid quantum communication and QIP. For instance, in a
hybrid QIP based on SPS qubits and CS qubits, entanglement transfer may be
necessary between SPS-qubit-based quantum processors and CS-qubit-based
quantum processors.

The new and rapidly growing field of circuit QED, consisting of microwave
radiation fields and fixed artificial atoms, offers extremely exciting
prospects for solid-state QIP [20-28]. In the following, we will present an
approach to transfer the entangled state (1) between $n$\ SPS qubits and $n$%
\ CS qubits, in a circuit QED system that consists of $2n$\ microwave
cavities coupled to a superconducting flux qutrit. In our proposal, the two
logic states $\left\vert i\right\rangle $ and $\left\vert \overline{i}%
\right\rangle $ of a SPS qubit are denoted as the vacuum state $\left\vert
0\right\rangle $ and the single-photon state $\left\vert 1\right\rangle $ of
a cavity; while the two logic states $\left\vert i\right\rangle $ and $%
\left\vert \overline{i}\right\rangle $ of a CS qubit are indicated by the
two coherent states $\left\vert \pm \alpha \right\rangle $ of a cavity,
which can be quasi-orthogonal for a sufficiently large $\alpha $ (for
instance, $\left\vert \left\langle \alpha \right\vert \left. -\alpha
\right\rangle \right\vert =e^{-2\left\vert \alpha \right\vert ^{2}}\sim
10^{-3}$ for $\alpha =1.86$). As is well known, coherent states are
eigenstates of the photon annihilation operator and thus tolerant to
single-photon loss [11].

In this work, we will explicitly show how to transfer the entangled state
(1) from $n$ SPS qubits onto $n$ CS qubits. Note that since no measurement
is involved, i.e., all basic operations are unitary, the entangled state (1)
can be transferred back from $n$ CS qubits onto $n$ SPS qubits by performing
reverse operations, in the case when the dissipation of the system is
negligible.

The present proposal has the following features: (i) Because only one
superconducting flux qutrit is used, the circuit complexity is reduced; (ii)
The state transfer time is independent of the number of qubits, thus it does
not increase with the increase of the number of qubits; (iii) The
higher-energy level of the qutrit is not excited during the entire
operation, thus the effect of decoherence from this level is suppressed; and
(iv) Since measurement is not required, the entangled state can be
transferred in a deterministic manner when the dissipation is negligible.
This proposal is universal and can be applied to accomplish the same task,
in a physical system which consists of $2n$\ microwave or optical cavities
coupled to a natural or artificial atom. To the best of our knowledge, how
to transfer quantum entangled states between SPS qubits and CS qubits has
not been reported yet.

We should point out that this proposal works by applying the rotating-wave
approximation (RWA). As is well known, the RWA is often used in quantum
optics and quantum information processing. As an open question, it would be
interesting to generalize the protocol to a more general case without
assuming the RWA, which usually applies in the ultrastrong coupling regime
(for a review, see [29]).

This paper is organized as follows. In Sec. II, we introduce a few types of
interaction and state evolution, which are used for transferring entangled
states. In Sec. III, we show how to transfer the entangled states of $n$\
SPS qubits onto $n$\ CS qubits. In Sec. IV, as an example, we investigate
the experimental feasibility for transferring a Bell state of two SPS qubits
onto two CS qubits in a circuit QED setup. A brief conclusion is given in
Sec. V.

\begin{figure}[tbp]
\begin{center}
\includegraphics[width=12.5 cm, clip]{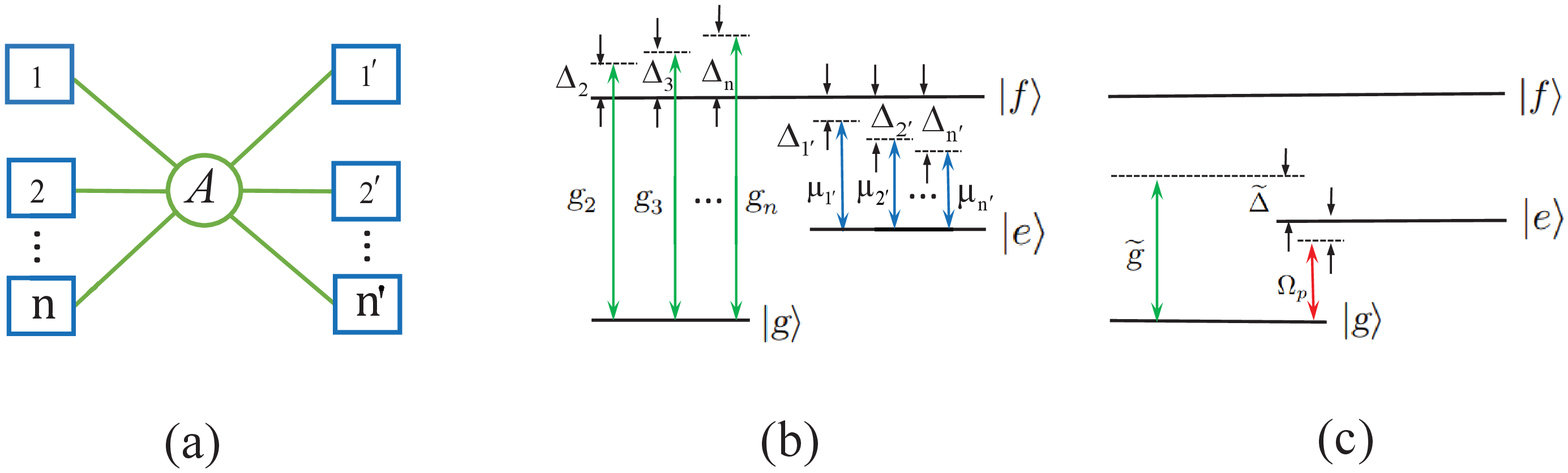} \vspace*{-0.08in}
\end{center}
\caption{(color online) (a) Diagram of $2n$ microwave cavities coupled to a
superconducting flux qutrit (the circle $A$ in the middle). Each square
represents a cavity, which can be a one- or three-dimensional cavity. The
qutrit is capacitively or inductively coupled to each cavity. (b)
Illustration of $n-1$ cavities $(2,3,...,n)$ off resonant with the $%
\left\vert g\right\rangle \leftrightarrow \left\vert f\right\rangle $
transition of the qutrit, while $n$ cavities $(1^{\prime },2^{\prime
},...,n^{\prime })$ off resonant with the $\left\vert e\right\rangle
\leftrightarrow \left\vert f\right\rangle $ transition of the qutrit. (c)
Illustration of cavity $1^{\prime }$ and a microwave pulse off resonant with
the $\left\vert g\right\rangle \leftrightarrow \left\vert e\right\rangle $
transition of the qutrit. }
\label{fig:1}
\end{figure}

\section{Type of interaction and state evolution}

Consider $n$\ single-mode cavities $(1,2,...,n)$\ and another $n$\
single-mode cavities $(1^{\prime },2^{\prime },...,n^{\prime })$. The $2n$\
cavities are connected by a superconducting flux qutrit $A$\ (Fig. 1a). The
three levels of the coupler qutrit are labelled as $\left\vert
g\right\rangle ,\left\vert e\right\rangle ,$and $\left\vert f\right\rangle $%
\ (Fig. 1b). In this section, we will introduce a few types of interaction,
the corresponditing Hamiltonians, and the state evolutions, which will be
used for the entangled state transfer in next section.

\subsection{Qutrit-cavity resonant interaction}

Consider cavity $1$\ on resonance with the $\left\vert g\right\rangle $\ $%
\leftrightarrow $\ $\left\vert e\right\rangle $\ transition of the qutrit.
In the interaction picture and after making the rotating-wave approximation
(RWA), the Hamiltonian is given by (hereafter assuming $\hbar =1$)\
\begin{equation}
H_{1}=g_{r}\hat{a}_{1}\sigma _{eg}^{+}+\text{h.c.},
\end{equation}%
where $\sigma _{eg}^{+}=\left\vert e\right\rangle \left\langle g\right\vert
, $ $g_{r}$\ is the coupling strength between cavity $1$\ and the $%
\left\vert g\right\rangle $\ $\leftrightarrow $\ $\left\vert e\right\rangle $%
\ transition, and $\hat{a}_{1}$\ is the photon annihilation operator of
cavity $1$. Under this Hamiltonian, it is easy to get the following state
evolution
\begin{eqnarray}
\left\vert e\right\rangle \left\vert 0\right\rangle _{1} &\rightarrow &\cos
g_{r}t\left\vert e\right\rangle \left\vert 0\right\rangle _{1}-i\sin
g_{r}t\left\vert g\right\rangle \left\vert 1\right\rangle _{1},  \notag \\
\left\vert g\right\rangle \left\vert 1\right\rangle _{1} &\rightarrow &\cos
g_{r}t\left\vert g\right\rangle \left\vert 1\right\rangle _{1}-i\sin
g_{r}t\left\vert e\right\rangle \left\vert 0\right\rangle _{1},
\end{eqnarray}%
while the state $\left\vert g\right\rangle \left\vert 0\right\rangle _{1}$\
remains unchanged.

\subsection{Off-resonant interaction between the coupler qutrit and the $%
\left( \mathbf{2n-1}\right) $ cavities}

Consider that the qutrit is decoupled from cavity $1$\ but interacts with
the $\left( 2n-1\right) $\ cavities ($2,3,...,n,1^{\prime },2^{\prime
},...,n^{\prime }$). Assume that cavity $j$\ is off resonant with the $%
\left\vert g\right\rangle \leftrightarrow \left\vert f\right\rangle $\
transition with coupling constant $g_{j}$\ and detuning $\Delta _{j}$\ $%
(j=2,3,...,n),$\ while highly detuned (decoupled) from other intra-level
transitions (Fig.~1b). In addition, suppose that cavity $j^{\prime }$\ is
off resonant with the $\left\vert e\right\rangle \leftrightarrow \left\vert
f\right\rangle $\ transition with coupling constant $\mu _{j^{\prime }}$\
and detuning $\Delta _{j^{\prime }}$\ $(j^{\prime }=1^{\prime },2^{\prime
},...,n^{\prime }),$\ while highly detuned (decoupled) from other
intra-level transitions (Fig.~1b). After these considerations, in the
interaction picture and after the RWA, the Hamiltonian is given by
\begin{equation}
H_{2}=\sum_{j=2}^{n}g_{j}\left( e^{i\Delta _{j}t}\hat{a}_{j}\sigma _{fg}^{+}+%
\text{h.c.}\right) +\sum_{j^{\prime }=1^{\prime }}^{n^{\prime }}\mu
_{j^{\prime }}\left( e^{i\Delta _{j^{\prime }}t}\hat{b}_{j^{\prime }}\sigma
_{fe}^{+}+\text{h.c.}\right) ,
\end{equation}%
where\ $\hat{a}_{j}$\ ($\hat{b}_{j^{\prime }}$) is the photon annihilation
operator of cavity $j$\ ($j^{\prime }$); $\sigma _{fg}^{+}=\left\vert
f\right\rangle \left\langle g\right\vert ,$\ $\sigma _{fe}^{+}=\left\vert
f\right\rangle \left\langle e\right\vert ,$\ $\Delta _{j}=\omega
_{fg}-\omega _{c_{j}}$\ and $\Delta _{j^{\prime }}=\omega _{fe}-\omega
_{c_{j^{\prime }}}$.$\ $Here, $\omega _{fg}$\ ($\omega _{fe}$) is the $%
\left\vert g\right\rangle \leftrightarrow \left\vert f\right\rangle $\ ($%
\left\vert e\right\rangle \leftrightarrow \left\vert f\right\rangle $)
transition frequency of the qutrit, while $\omega _{c_{j}}$\ ($\omega
_{c_{j^{\prime }}}$) is the frequency of cavity $j$\ (cavity $j^{\prime }$).

Under the large detunings $\left\vert \Delta _{j}\right\vert \gg g_{j}$\ and$%
\;\left\vert \Delta _{j^{\prime }}\right\vert \gg \mu _{j^{\prime },}$\ no
energy exchange is induced between the coupler qutrit and the cavities. When
the following conditions satisfy, i.e.,
\begin{equation}
\frac{\left\vert \Delta _{j}-\Delta _{k}\right\vert }{\left\vert \Delta
_{j}^{-1}\right\vert +\left\vert \Delta _{k}^{-1}\right\vert }\gg g_{j}g_{k},%
\frac{\left\vert \Delta _{j^{\prime }}-\Delta _{k^{\prime }}\right\vert }{%
\left\vert \Delta _{j^{\prime }}^{-1}\right\vert +\left\vert \Delta
_{k^{\prime }}^{-1}\right\vert }\gg \mu _{j^{\prime }}\mu _{k^{\prime }},%
\frac{\left\vert \Delta _{j}-\Delta _{k^{\prime }}\right\vert }{\left\vert
\Delta _{j}^{-1}\right\vert +\left\vert \Delta _{k^{\prime
}}^{-1}\right\vert }\gg g_{j}\mu _{k^{\prime }},
\end{equation}%
(where $j,k\in \{2,3,...,n\},$\ $j^{\prime },k^{\prime }\in \{1^{\prime
},2^{\prime },...,n^{\prime }\},$\ $j\neq k,$\ $j^{\prime }\neq k^{\prime }$%
), no interaction between cavities ($2,3,...,n,1^{\prime },2^{\prime
},...,n^{\prime }$) is induced by the coupler qutrit. Thus, from the
Hamiltonian (4), one can obtain the following effective Hamiltonian [30,31]
\begin{eqnarray}
H_{\mathrm{e}} &=&-\sum_{j=2}^{n}\frac{g_{j}^{2}}{\Delta _{j}}\left( \sigma
_{g}\hat{a}_{j}^{+}\hat{a}_{j}-\sigma _{f}\hat{a}_{j}\hat{a}_{j}^{+}\right)
\notag \\
&&\ -\sum_{j^{\prime }=1^{\prime }}^{n^{\prime }}\frac{\mu _{j^{\prime }}^{2}%
}{\Delta _{j^{\prime }}}\left( \sigma _{e}\hat{b}_{j^{\prime }}^{+}\hat{b}%
_{j^{\prime }}\hat{-}\sigma _{f}\hat{b}_{j^{\prime }}\hat{b}_{j^{\prime
}}^{+}\right) ,
\end{eqnarray}%
where $\sigma _{g}=\left\vert g\right\rangle \left\langle g\right\vert
,\sigma _{e}=\left\vert e\right\rangle \left\langle e\right\vert ,\sigma
_{f}=\left\vert f\right\rangle \left\langle f\right\vert .$ In Eq. (6), the
terms in the first line represent the ac-Stark shifts of the levels $%
\left\vert g\right\rangle $\ and $\left\vert f\right\rangle $\ induced by
the cavities $(2,3,...,n)$, while the terms in the second line are the
ac-Stark shifts of the levels $\left\vert e\right\rangle $\ and $\left\vert
f\right\rangle $\ caused by the cavities $(1^{\prime },2^{\prime
},...,n^{\prime })$.

In the case of the level $\left\vert f\right\rangle $\ being not occupied,
the Hamiltonian (6) reduces to
\begin{equation}
H_{\mathrm{e}}=-\sum_{j=2}^{n}\lambda _{j}\sigma _{g}\hat{a}_{j}^{+}\hat{a}%
_{j}-\sum_{j^{\prime }=1^{\prime }}^{n^{\prime }}\lambda _{j^{\prime
}}\sigma _{e}\hat{b}_{j^{\prime }}^{+}\hat{b}_{j^{\prime }},
\end{equation}%
where $\lambda _{j}=g_{j}^{2}/\Delta _{j}$ and $\lambda _{j^{\prime }}=\mu
_{j^{\prime }}^{2}/\Delta _{j^{\prime }}.$ Under this Hamiltonian (7), it is
straightforward to obtain%
\begin{eqnarray}
\left\vert g\right\rangle \left\vert 0\right\rangle _{j} &\rightarrow
&\left\vert g\right\rangle \left\vert 0\right\rangle _{j},\text{ }\left\vert
g\right\rangle \left\vert 1\right\rangle _{j}\rightarrow e^{i\lambda
_{j}t}\left\vert g\right\rangle \left\vert 1\right\rangle _{j},\text{\ }%
\left\vert g\right\rangle \left\vert \alpha \right\rangle _{j^{\prime
}}\rightarrow \left\vert g\right\rangle \left\vert \alpha \right\rangle
_{j^{\prime }},\text{\ \ }  \notag \\
\left\vert e\right\rangle \left\vert 0\right\rangle _{j} &\rightarrow
&\left\vert e\right\rangle \left\vert 0\right\rangle _{j},\text{ }\left\vert
e\right\rangle \left\vert 1\right\rangle _{j}\rightarrow \left\vert
e\right\rangle \left\vert 1\right\rangle _{j},\text{ }\left\vert
e\right\rangle \left\vert \alpha \right\rangle _{j^{\prime }}\rightarrow
\left\vert e\right\rangle \left\vert e^{i\lambda _{j^{\prime }}t}\alpha
\right\rangle _{j^{\prime }},
\end{eqnarray}%
where $j\in \left\{ 2,3,...,n\right\} $ and $j^{\prime }\in \{1^{\prime
},2^{\prime },...,n^{\prime }\}.$

\subsection{Driving the qutrit conditional on the cavity's state}

Consider cavity $1^{\prime }$\ to be off resonant with the $\left\vert
g\right\rangle $\ $\leftrightarrow $\ $\left\vert e\right\rangle $\
transition of the qutrit, with coupling strength $\widetilde{g}$\ and
detuning $\widetilde{\Delta }=\omega _{eg}-\widetilde{\omega }_{c_{1^{\prime
}}}$\ (Fig. 1c). Here, $\widetilde{\omega }_{c_{1^{\prime }}}$\ is the
adjusted frequency of cavity $1^{\prime }$. Meanwhile, assume that a
classical pulse of frequency $\omega _{p}$\ and initial phase $\phi $ is
applied to the qutrit$.$\

Under the above assumptions, the Hamiltonian of the system in the
interaction picture and after the RWA is given by%
\begin{equation}
H_{3}=\widetilde{g}e^{i\widetilde{\Delta }t}\hat{a}_{1^{\prime }}\sigma
_{eg}^{+}+\Omega _{p}e^{-i\left[ \left( \omega _{p}-\omega _{eg}\right)
t+\phi \right] }\sigma _{eg}^{+}+\text{h.c.,}
\end{equation}%
\ where $\Omega _{p}$ is the Rabi frequency of the pulse.

Under the large detuning condition $\left\vert \widetilde{\Delta }%
\right\vert \gg \tilde{g},$\ and assuming $\left\vert \widetilde{\Delta }%
\right\vert \gg \Omega _{p},$\ we can obtain the following effective
Hamiltonian [22,31]
\begin{equation}
H_{\mathrm{e}}=\frac{\widetilde{g}^{2}}{\tilde{\Delta}}\left( \hat{a}%
_{1^{\prime }}^{+}\hat{a}_{1^{\prime }}+\frac{1}{2}\right) \sigma
_{z}+\left( \Omega _{p}e^{-i\left[ \left( \omega _{p}-\omega _{eg}\right)
t+\phi \right] }\sigma _{eg}^{+}+\text{h.c.}\right) ,
\end{equation}%
where $\sigma _{z}=\left\vert e\right\rangle \left\langle e\right\vert
-\left\vert g\right\rangle \left\langle g\right\vert .$\ Here, the
stark-shift effect of the qubit induced by the pulse is negligible because
of $\left\vert \widetilde{\Delta }\right\vert \gg \Omega _{p}.$ In a
rotating frame under the Hamiltonian $H_{0}=\frac{\widetilde{g}^{2}}{\tilde{%
\Delta}}\left( \hat{a}_{1^{\prime }}^{+}\hat{a}_{1^{\prime }}+\frac{1}{2}%
\right) \sigma _{z},$\ and for $\omega _{p}=\omega _{eg}+\widetilde{g}^{2}/%
\widetilde{\Delta },$\ one obtains
\begin{equation}
H_{\mathrm{e}}=\Omega _{p}e^{-i\phi }e^{i4\widetilde{\omega }\hat{a}%
_{1^{\prime }}^{+}\hat{a}_{1^{\prime }}t}\sigma _{eg}^{+}+\text{h.c.,}
\end{equation}%
where $\widetilde{\omega }=\widetilde{g}^{2}/\left( 2\widetilde{\Delta }%
\right) .$ When the cavity is in the vacuum state, one can easily find that
this Hamiltonian reduce to $H_{\mathrm{eff}}=\Omega _{p}e^{-i\phi }\sigma
_{eg}^{+}+$ H.c., which rotates the qubit's state as follows%
\begin{eqnarray}
\left\vert g\right\rangle \left\vert 0\right\rangle _{1^{\prime }}
&\rightarrow &\left( \cos \Omega _{p}t\left\vert g\right\rangle -ie^{-i\phi
}\sin \Omega _{p}t\left\vert e\right\rangle \right) \left\vert
0\right\rangle _{1^{\prime }},  \notag \\
\left\vert e\right\rangle \left\vert 0\right\rangle _{1^{\prime }}
&\rightarrow &\left( -ie^{i\phi }\sin \Omega _{p}t\left\vert g\right\rangle
+\cos \Omega _{p}t\left\vert e\right\rangle \right) \left\vert
0\right\rangle _{1^{\prime }}.
\end{eqnarray}%
On the other hand, when the cavity is in the coherent state $\left\vert
2\alpha \right\rangle ,$\ the Hamiltonian (11) becomes $H_{\mathrm{eff}%
}=\Omega _{p}e^{-i\phi }e^{i4\widetilde{\omega }\overline{n}t}\sigma
_{eg}^{+}+$h.c., with $\overline{n}=4\left\vert \alpha \right\vert ^{2}.$\
For a sufficient large $\alpha $, such that $4\left\vert \widetilde{\omega }%
\right\vert \overline{n}\gg \Omega _{p}$\ (large detuning), it is easy to
see that the qubit's state is not changed by the driving pulse, i.e.,
\begin{eqnarray}
\left\vert g\right\rangle \left\vert 2\alpha \right\rangle _{1^{\prime }}
&\rightarrow &\left\vert g\right\rangle \left\vert 2\alpha \right\rangle
_{1^{\prime }},  \notag \\
\left\vert e\right\rangle \left\vert 2\alpha \right\rangle _{1^{\prime }}
&\rightarrow &\left\vert e\right\rangle \left\vert 2\alpha \right\rangle
_{1^{\prime }}.
\end{eqnarray}%
After returning to the original interaction picture by performing a unitary
operation $U=e^{-iH_{0}},$\ we have from Eqs. (12) and (13)%
\begin{eqnarray}
\left\vert g\right\rangle \left\vert 0\right\rangle _{1^{\prime }}
&\rightarrow &\left( e^{i\widetilde{\omega }t}\cos \Omega _{p}t\left\vert
g\right\rangle -ie^{-i\phi }e^{-i\widetilde{\omega }t}\sin \Omega
_{p}t\left\vert e\right\rangle \right) \left\vert 0\right\rangle _{1^{\prime
}},  \notag \\
\left\vert e\right\rangle \left\vert 0\right\rangle _{1^{\prime }}
&\rightarrow &\left( -ie^{i\phi }e^{i\widetilde{\omega }t}\sin \Omega
_{p}t\left\vert g\right\rangle +e^{-i\widetilde{\omega }t}\cos \Omega
_{p}t\left\vert e\right\rangle \right) \left\vert 0\right\rangle _{1^{\prime
}},
\end{eqnarray}%
and
\begin{eqnarray}
\left\vert g\right\rangle \left\vert 2\alpha \right\rangle _{1^{\prime }}
&\rightarrow &e^{i\widetilde{\omega }t}\left\vert g\right\rangle \left\vert
e^{i2\widetilde{\omega }t}2\alpha \right\rangle _{1^{\prime }},  \notag \\
\left\vert e\right\rangle \left\vert 2\alpha \right\rangle _{1^{\prime }}
&\rightarrow &e^{-i\widetilde{\omega }t}\left\vert e\right\rangle \left\vert
e^{-i2\widetilde{\omega }t}2\alpha \right\rangle _{1^{\prime }}.
\end{eqnarray}%
In the next section, we will show how to use the results (3), (8), (14) and
(15) to transfer entangled states from $n$\ PO qubits to $n$\ WO qubits.

\section{Transfer of SPS-qubit entangled states onto CS-qubits}

The setup is shown in Fig. 1(a). Initially, the coupler qutrit is in the
ground state $\left\vert g\right\rangle $\ and decoupled from the cavity
system. The $n$\ SPS qubits are initially in the entangled state
\begin{equation}
c\left\vert 0\right\rangle _{1}\left\vert 0\right\rangle _{2}...\left\vert
0\right\rangle _{n}+d\left\vert 1\right\rangle _{1}\left\vert 1\right\rangle
_{2}...\left\vert 1\right\rangle _{n},
\end{equation}%
where $\left\vert 0\right\rangle $\ and $\left\vert 1\right\rangle $\ are
the vacuum state and the single-photon state.

Suppose that each of the $n$\ cavities ($1^{\prime },2^{\prime
},...,n^{\prime }$) is in a coherent state $\left\vert \alpha \right\rangle $%
. Thus, the initial state of the whole system is given by
\begin{equation}
\left( c\left\vert 0\right\rangle _{1}\left\vert 0\right\rangle
_{2}...\left\vert 0\right\rangle _{n}+d\left\vert 1\right\rangle
_{1}\left\vert 1\right\rangle _{2}...\left\vert 1\right\rangle _{n}\right)
\left\vert \alpha \right\rangle _{1^{\prime }}\left\vert \alpha
\right\rangle _{2^{\prime }}...\left\vert \alpha \right\rangle _{n^{\prime
}}\left\vert g\right\rangle ,
\end{equation}%
where subscripts $1,2,...,n$\ represent cavities $1,2,...,n$\ and subscripts
$1^{\prime },2^{\prime },...,n^{\prime }$\ represent cavities $1^{\prime
},2^{\prime },...,n^{\prime },$\ respectively. The procedure for
transferring the entangled state is listed below:

Step (i): The purpose of this step is to convert the states $\left\vert
0\right\rangle $\ and $\left\vert 1\right\rangle $\ of each of cavities ($%
2,3,...,n$) into $\left\vert +\right\rangle $\ and $\left\vert
-\right\rangle ,$\ respectively. Here, $\left\vert \pm \right\rangle =\left(
\left\vert 0\right\rangle \pm \left\vert 1\right\rangle \right) /\sqrt{2}.$
Note that this state convertion can be easily implemented by performing
local resonant operations on cavity $j$\ and an auxilliary qubit $j$\
(initially in the ground state) placed in cavity $j,$\ i.e., applying
resonant interaction of qubit $j$\ with cavity $j$\ and resonant interaction
of qubit $j$ with a classical pulse [32]. After this step of operation, the
state (17) becomes
\begin{equation}
\left( c\left\vert 0\right\rangle _{1}\left\vert +\right\rangle
_{2}\left\vert +\right\rangle _{3}...\left\vert +\right\rangle
_{n}+d\left\vert 1\right\rangle _{1}\left\vert -\right\rangle _{2}\left\vert
-\right\rangle _{3}...\left\vert -\right\rangle _{n}\right) \left\vert
\alpha \right\rangle _{1^{\prime }}\left\vert \alpha \right\rangle
_{2^{\prime }}...\left\vert \alpha \right\rangle _{n^{\prime }}\left\vert
g\right\rangle .
\end{equation}

Step (ii):\ Adjust the level spacings of the coupler qutrit to have cavity $%
1 $\ on resonance with the $\left\vert g\right\rangle \leftrightarrow
\left\vert e\right\rangle $\ transition. According to Eq. (3), one has $%
\left\vert 0\right\rangle _{1}\left\vert g\right\rangle \rightarrow
\left\vert 0\right\rangle _{1}\left\vert g\right\rangle $\ and $\left\vert
1\right\rangle _{1}\left\vert g\right\rangle \rightarrow -i\left\vert
0\right\rangle _{1}\left\vert e\right\rangle $\ for an interaction time $%
t=\pi /\left( 2g_{r}\right) .$\ Hence, the state (18) becomes
\begin{equation}
\left\vert 0\right\rangle _{1}\left( c\left\vert g\right\rangle \left\vert
+\right\rangle _{2}\left\vert +\right\rangle _{3}...\left\vert
+\right\rangle _{n}-id\left\vert e\right\rangle \left\vert -\right\rangle
_{2}\left\vert -\right\rangle _{3}...\left\vert -\right\rangle _{n}\right)
\left\vert \alpha \right\rangle _{1^{\prime }}\left\vert \alpha
\right\rangle _{2^{\prime }}...\left\vert \alpha \right\rangle _{n^{\prime
}}.
\end{equation}%
Then, the level spacings of the coupler qutrit is adjusted back such that
the qutrit is decoupled from cavity $1.$

Step (iii): Adjust the level spacings of the coupler qutrit so that the
qutrit is off resonant with the cavities ($2,3,...,n,1^{\prime },2^{\prime
},...,n^{\prime }$) (Fig. 1b) to achieve an effective Hamiltonian (7).
According to Eq. (8), the state (19) changes to
\begin{eqnarray}
&&\left\vert 0\right\rangle _{1}\left[ c\left\vert g\right\rangle \left(
\left\vert 0\right\rangle _{2}+e^{i\lambda _{2}t}\left\vert 1\right\rangle
_{2}\right) \left( \left\vert 0\right\rangle _{3}+e^{i\lambda
_{3}t}\left\vert 1\right\rangle _{3}\right) ...\left( \left\vert
0\right\rangle _{n}+e^{i\lambda _{n}t}\left\vert 1\right\rangle _{n}\right)
\left\vert \alpha \right\rangle _{1^{\prime }}\left\vert \alpha
\right\rangle _{2^{\prime }}..\left\vert \alpha \right\rangle _{n^{\prime
}}\right.  \notag \\
&&-id\left. \left\vert e\right\rangle \left\vert -\right\rangle
_{2}\left\vert -\right\rangle _{3}...\left\vert -\right\rangle
_{n}\left\vert e^{i\lambda _{1^{\prime }}t}\alpha \right\rangle _{1^{\prime
}}\left\vert e^{i\lambda _{2^{\prime }}t}\alpha \right\rangle _{2^{\prime
}}..\left\vert e^{i\lambda _{n^{\prime }}t}\alpha \right\rangle _{n^{\prime
}}\right] .
\end{eqnarray}%
Now let us set
\begin{equation}
\left\vert \lambda _{2}\right\vert =\left\vert \lambda _{3}\right\vert
...=\left\vert \lambda _{n}\right\vert =\lambda ,\text{ }\left\vert \lambda
_{1^{\prime }}\right\vert =\left\vert \lambda _{2^{\prime }}\right\vert
...=\left\vert \lambda _{n^{\prime }}\right\vert =\lambda .
\end{equation}%
Thus, for an interaction time $t=\pi /\lambda ,$\ the state (20) becomes
\begin{equation}
\left\vert 0\right\rangle _{1}\left\vert -\right\rangle _{2}\left\vert
-\right\rangle _{3}...\left\vert -\right\rangle _{n}\left( c\left\vert
g\right\rangle \left\vert \alpha \right\rangle _{1^{\prime }}\left\vert
\alpha \right\rangle _{2^{\prime }}..\left\vert \alpha \right\rangle
_{n^{\prime }}-id\left\vert e\right\rangle \left\vert -\alpha \right\rangle
_{1^{\prime }}\left\vert -\alpha \right\rangle _{2^{\prime }}..\left\vert
-\alpha \right\rangle _{n^{\prime }}\right) .
\end{equation}%
Then, the level spacings of the qutrit is adjusted back such that the qutrit
is decoupled from the $2n$\ cavities.

Step (iv): Apply a classical driving field on resonance with cavity $%
1^{\prime }$\ to obtain a displacement operator $D\left( \alpha \right) $ in
phase space, which results in the transformation $\left\vert \alpha
\right\rangle _{1^{\prime }}\rightarrow \left\vert 2\alpha \right\rangle
_{1^{\prime }}$\ and $\left\vert -\alpha \right\rangle _{1^{\prime
}}\rightarrow \left\vert 0\right\rangle _{1^{\prime }}$. Thus, we have from
Eq. (22)
\begin{equation}
\left\vert 0\right\rangle _{1}\left\vert -\right\rangle _{2}\left\vert
-\right\rangle _{3}...\left\vert -\right\rangle _{n}\left( c\left\vert
g\right\rangle \left\vert 2\alpha \right\rangle _{1^{\prime }}\left\vert
\alpha \right\rangle _{2^{\prime }}..\left\vert \alpha \right\rangle
_{n^{\prime }}-id\left\vert e\right\rangle \left\vert 0\right\rangle
_{1^{\prime }}\left\vert -\alpha \right\rangle _{2^{\prime }}..\left\vert
-\alpha \right\rangle _{n^{\prime }}\right) .
\end{equation}%
Step (v): Adjust the frequency of cavity $1^{\prime }$ to have it off
resonant with the $\left\vert g\right\rangle \leftrightarrow \left\vert
e\right\rangle $ transition of the qutrit. Meanwhile apply a classical pulse
(with initial phase $\phi =\pi $\ and frequency $\omega _{p}=\omega _{eg}+%
\widetilde{g}^{2}/\widetilde{\Delta }$)\ to the qutrit.\ According to Eqs.
(14) and (15), for $t=\pi /\left( 2\Omega _{p}\right) ,$\ we have $%
\left\vert e\right\rangle \left\vert 0\right\rangle _{1^{\prime
}}\rightarrow ie^{i\widetilde{\omega }t}\left\vert g\right\rangle \left\vert
0\right\rangle _{1^{\prime }}$\ while the state $\left\vert g\right\rangle
\left\vert 2\alpha \right\rangle _{1^{\prime }}$ becomes $e^{i\widetilde{%
\omega }t}\left\vert g\right\rangle \left\vert e^{i2\widetilde{\omega }%
t}2\alpha \right\rangle _{1^{\prime }}$. Thus, we have from Eq. (23)
\begin{equation}
e^{i\widetilde{\omega }t}\left\vert g\right\rangle \left\vert 0\right\rangle
_{1}\left\vert -\right\rangle _{2}\left\vert -\right\rangle
_{3}...\left\vert -\right\rangle _{n}\left\{ c\left\vert e^{i2\widetilde{%
\omega }t}2\alpha \right\rangle _{1^{\prime }}\left\vert \alpha
\right\rangle _{2^{\prime }}..\left\vert \alpha \right\rangle _{n^{\prime
}}+d\left\vert 0\right\rangle _{1^{\prime }}\left\vert -\alpha \right\rangle
_{2^{\prime }}..\left\vert -\alpha \right\rangle _{n^{\prime }}\right\} .
\end{equation}%
If the operation time of this step satisfies $2\widetilde{\omega }t=2m\pi $ (%
$m$ is an integer), we have $e^{i2\widetilde{\omega }t}=1.$ Thus, the state
(24) becomes
\begin{equation}
\left\vert g\right\rangle \left\vert 0\right\rangle _{1}\left\vert
-\right\rangle _{2}\left\vert -\right\rangle _{3}...\left\vert
-\right\rangle _{n}\left( c\left\vert 2\alpha \right\rangle _{1^{\prime
}}\left\vert \alpha \right\rangle _{2^{\prime }}..\left\vert \alpha
\right\rangle _{n^{\prime }}+d\left\vert 0\right\rangle _{1^{\prime
}}\left\vert -\alpha \right\rangle _{2^{\prime }}..\left\vert -\alpha
\right\rangle _{n^{\prime }}\right) ,
\end{equation}%
where we have dropped off the global phase $e^{i\widetilde{\omega }t}$.
After this step of operation, the frequency of cavity $1^{\prime }$\ is
adjusted such that cavity $1^{\prime }$ is decoupled from the qutrit.

Step (vi): Apply a classical driving field on resonance with cavity $%
1^{\prime },$\ to obtain a displacement operator $D\left( -\alpha \right) $\
resulting in the transformation $\left\vert 2\alpha \right\rangle
_{1^{\prime }}\rightarrow \left\vert \alpha \right\rangle _{1^{\prime }}$\
and $\left\vert 0\right\rangle _{1^{\prime }}\rightarrow \left\vert -\alpha
\right\rangle _{1^{\prime }}$. Hence, we have from Eq. (25)
\begin{equation}
\left\vert g\right\rangle \left\vert 0\right\rangle _{1}\left\vert
-\right\rangle _{2}\left\vert -\right\rangle _{3}...\left\vert
-\right\rangle _{n}\otimes \left( c\left\vert \alpha \right\rangle
_{1^{\prime }}\left\vert \alpha \right\rangle _{2^{\prime }}..\left\vert
\alpha \right\rangle _{n^{\prime }}+d\left\vert -\alpha \right\rangle
_{1^{\prime }}\left\vert -\alpha \right\rangle _{2^{\prime }}..\left\vert
-\alpha \right\rangle _{n^{\prime }}\right) ,
\end{equation}%
which shows that the $n$\ CS qubits ($1^{\prime },2^{\prime },...,n^{\prime
} $) are in the entangled state $c\left\vert \alpha \right\rangle
_{1^{\prime }}\left\vert \alpha \right\rangle _{2^{\prime }}..\left\vert
\alpha \right\rangle _{n^{\prime }}+d\left\vert -\alpha \right\rangle
_{1^{\prime }}\left\vert -\alpha \right\rangle _{2^{\prime }}..\left\vert
-\alpha \right\rangle _{n^{\prime }},$\ i.e., the original entangled state
(16) of the $n$\ SPS qubits has been transferred onto the $n$\ CS qubits
after the above operations. Here, note that in Eq. (26), the two logic
states of CS qubit $j^{\prime }$\ ($j^{\prime }=1^{\prime },2^{\prime
},...,n^{\prime }$) are encoded with two coherent states $\left\vert \alpha
\right\rangle $\ and $\left\vert -\alpha \right\rangle $\ of cavity $%
j^{\prime }.$\

To ensure that each step described above is performed by unitary
operations, the two wave vectors in each bracket of
Eqs. (24-26) should be orthogonal. Note that this requirement can be
met for a large enough $\alpha .$ The reason for this is as follows. 
Because of $\left\vert \left\langle \alpha
\right\vert \left. -\alpha \right\rangle \right\vert =$ $\left\vert
\left\langle 2\alpha \right\vert \left. 0\right\rangle \right\vert
=e^{-2\left\vert \alpha \right\vert ^{2}},$ the two coherent states $%
\left\vert \alpha \right\rangle $\ and $\left\vert -\alpha \right\rangle $
or the two states $\left\vert 0\right\rangle $ and $\left\vert 2\alpha
\right\rangle $ involved in Eqs. (24-26) can be quasi-orthogonal for a
sufficiently large $\alpha ,$ as mentioned in the introduction.

Several other points need to be made as follows:

(i) As shown above, there is no need of measurement for the state transfer,
i.e., all basic operations are unitary. Thus, by performing reverse unitary
operations, the entangled state $c\left\vert \alpha \right\rangle
_{1^{\prime }}\left\vert \alpha \right\rangle _{2^{\prime }}..\left\vert
\alpha \right\rangle _{n^{\prime }}+d\left\vert -\alpha \right\rangle
_{1^{\prime }}\left\vert -\alpha \right\rangle _{2^{\prime }}..\left\vert
-\alpha \right\rangle _{n^{\prime }}$\ of $n$\ CS qubits can be converted
back into the entangled state $c\left\vert 0\right\rangle _{1}\left\vert
0\right\rangle _{2}...\left\vert 0\right\rangle _{n}+d\left\vert
1\right\rangle _{1}\left\vert 1\right\rangle _{2}...\left\vert
1\right\rangle _{n}$\ of $n$\ SPS qubits when the dissipation of the system
is negligible.

(ii) The total operation time for the state transfer is
\begin{equation}
t_{\mathrm{op}}=\pi /\lambda +\pi /\left( 2g_{r}\right) +\pi /\left( 2\Omega
_{p}\right) +\tau _{p}+2\tau _{\alpha }+4\tau _{d}+2\tau _{c},
\end{equation}%
which does not depend on the number of qubits involved in the entangled
state. Here, $\tau _{p}$ is the typical time for converting the states $%
\left\vert 0\right\rangle $\ and $\left\vert 1\right\rangle $\ of each of
cavities ($2,3,...,n$) into $\left\vert \pm \right\rangle $, $\tau _{\alpha
} $\ is the typical time for performing a displacement operation on the
cavity's coherent state, $\tau _{d}$\ is the typical time required for
adjusting the level spacings of the coupler qutrit, and $\tau _{c}$\ is the
typical time for adjusting the cavity frequency.

(iii) The state transfer requires adjusting the qutrit's level spacings to
have the qutrit coupled with (or decoupled from) the cavities. For
superconducting devices, their level spacings can be rapidly (within 1--3
ns) adjusted by varying external control parameters [33,34]. In addition, as
shown above, the frequency of cavity $1^{\prime }$\ needs to be adjusted to
have cavity $1^{\prime }$\ coupled with (or decoupled from) the qutrit. Note
that for superconducting microwave cavities, the cavity frequencies can be
fast (within a few nanoseconds)\ tuned in experiments [35,36].

(iv) Because of $\lambda _{j}=g_{j}^{2}/\Delta _{j}$\ and $\lambda
_{j^{\prime }}=\mu _{j^{\prime }}^{2}/\Delta _{j^{\prime }},$\ the condition
(21) turns out into
\begin{equation}
\frac{g_{2}^{2}}{\left\vert \Delta _{2}\right\vert }=\frac{g_{3}^{2}}{%
\left\vert \Delta _{3}\right\vert }=...=\frac{g_{n}^{2}}{\left\vert \Delta
_{n}\right\vert }=\frac{\mu _{1^{\prime }}^{2}}{\left\vert \Delta
_{1^{\prime }}\right\vert }=\frac{\mu _{2^{\prime }}^{2}}{\left\vert \Delta
_{2^{\prime }}\right\vert }=\frac{\mu _{3^{\prime }}^{2}}{\left\vert \Delta
_{3^{\prime }}\right\vert }=...=\frac{\mu _{n^{\prime }}^{2}}{\left\vert
\Delta _{n^{\prime }}\right\vert }.
\end{equation}%
Due to $\Delta _{j}=\omega _{fg}-\omega _{c_{j}}$\ $\left( 2,3,...,n\right) $
and $\Delta _{j^{\prime }}=\omega _{fe}-\omega _{c_{j^{\prime }}}$ $\left(
1^{\prime },2^{\prime },...,n^{\prime }\right) ,$\ the equality given in Eq.
(28)\ can be easily established by carefully selecting $\Delta _{j}$\ (via
adjusting the frequency $\omega _{c_{j}}$\ of cavity $j$) and carefully
selecting $\Delta _{j^{\prime }}$\ (via adjusting the frequency $\omega
_{c_{j^{\prime }}}$\ of cavity $j^{\prime }$).

(v) As described above, for step (v), the operation time should satify both $%
t=\pi /\left( 2\Omega _{p}\right) $\ and $2\widetilde{\omega }t=2m\pi ,$\
i.e.,
\begin{equation}
\Omega _{p}=\frac{\widetilde{g}^{2}}{4m\widetilde{\Delta }},
\end{equation}%
which can be easily met by adjusting the Rabi frequency of the pulse.

The present proposal works out for either $c=d$ or $c\neq d$. Note that the
case of $c=d$ is specially interesting, because Bell states and GHZs are of
great importance in quantum communication and QIP. It should be mentioned
that how to directly prepare the initial entangled state (16) with $c=d$\
was previously reported [37]. Alternatively, the state (16) with $c=d$ can
be created by preparing $n$ natural or artificial atoms (distributed in $n$
cavities) in a maximally entangled state $\left\vert g\right\rangle
_{1}\left\vert g\right\rangle _{2}...\left\vert g\right\rangle
_{n}+\left\vert e\right\rangle _{1}\left\vert e\right\rangle
_{2}...\left\vert e\right\rangle _{n}$ [38-42] and then having each atom (in
the excited state $\left\vert e\right\rangle $) emitting a single photon
into its cavity through local operations on each atom and its cavity.

\section{Possible experimental implementation}

\begin{figure}[tbp]
\begin{center}
\includegraphics[bb=24 178 830 421, width=12.5 cm, clip]{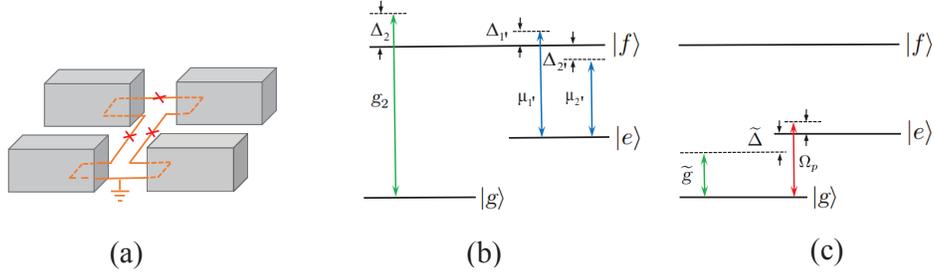} \vspace*{%
-0.08in}
\end{center}
\caption{(color online) (a) Schematic circuit of four 3D microwave cavities
coupled to a superconducting flux qutrit. The qutrit consists of three
Josephson junctions and a superconducting loop. (b) Illustration of cavity $%
2 $ off resonant with the $\left\vert g\right\rangle \leftrightarrow
\left\vert f\right\rangle $ transition of the qutrit, while cavities $%
1^{\prime }$ and $2^{\prime }$ off resonant with the $\left\vert
e\right\rangle \leftrightarrow \left\vert f\right\rangle $ transition of the
qutrit. (c) Illustration of cavity $1^{\prime }$ and a microwave pulse off
resonant with the $\left\vert g\right\rangle \leftrightarrow \left\vert
e\right\rangle $ transition of the qutrit. The unwanted couplings or
interactions of the qutrit's energy-level transitions with the cavities or
the pulse are not illustrated in (b) and (c) for simplicity, which however
are considered in our numerical simulation, as described in the Hamiltonians
(30) and (32).}
\label{fig:2}
\end{figure}

This section is aimed at investigating the experimental feasibility for
transferring a Bell state of two SPS qubits $(1,2)$\ onto two CS qubits $%
\left( 1^{\prime },2^{\prime }\right) $, by considering four 3D microwave
cavities coupled to a superconducting flux qutrit (Fig. 2a). In the
following, we will briefly discuss the fidelity for the Bell state transfer,
which is performed in a more realistic situation. Since steps (i), (ii),
(iv) and (vi) employ only the resonant interactions, these steps can be
completed within a very short time. In this sense, the influence of the
qutrit decoherence, the cavity decay, and the inter-cavity crosstalk is
negligibly small for steps (i), (ii), (iv) and (vi). Therefore, the
imperformance of the Bell state transfer is mainly caused by the system
dissipation and the inter-cavity crosstalk during the operation of steps
(iii) and (v), because these steps require reletively long operation times
due to the use of the qutrit-cavity off-resonant interaction.

After taking into account the unwanted couplings between the qutrit's level
transitions and the cavities as well as the inter-cavity crosstalk, the
Hamiltonian $H_{2}$\ in Eq. (4), with $n=2$\ and $n^{\prime }=2^{\prime }$\
for the present case, is modified as%
\begin{eqnarray}
H_{2}^{\prime } &=&\left( g_{2}e^{i\Delta _{2}t}\hat{a}_{2}\sigma
_{fg}^{+}+g_{2}^{\prime }e^{i\Delta _{2}^{\prime }t}\hat{a}_{2}\sigma
_{fe}^{+}+g_{2}^{\prime \prime }e^{i\Delta _{2}^{\prime \prime }t}\hat{a}%
_{2}\sigma _{eg}^{+}+\text{h.c.}\right)  \notag \\
&&+\left( \mu _{1^{\prime }}e^{i\Delta _{1^{\prime }}t}\hat{b}_{1^{\prime
}}\sigma _{fe}^{+}+\mu _{1^{\prime }}^{\prime }e^{i\Delta _{1^{\prime
}}^{\prime }t}\hat{b}_{1^{\prime }}\sigma _{fg}^{+}+\mu _{1^{\prime
}}^{\prime \prime }e^{i\Delta _{1^{\prime }}^{\prime \prime }t}\hat{b}%
_{1^{\prime }}\sigma _{eg}^{+}+\text{h.c.}\right)  \notag \\
&&+\left( \mu _{2^{\prime }}e^{i\Delta _{2^{\prime }}t}\hat{b}_{2^{\prime
}}\sigma _{fe}^{+}+\mu _{2^{\prime }}^{\prime }e^{i\Delta _{2^{\prime
}}^{\prime }t}\hat{b}_{2^{\prime }}\sigma _{fg}^{+}+\mu _{2^{\prime
}}^{\prime \prime }e^{i\Delta _{2^{\prime }}^{\prime \prime }t}\hat{b}%
_{2^{\prime }}\sigma _{eg}^{+}+\text{h.c.}\right)  \notag \\
&&+\varepsilon ,
\end{eqnarray}%
where the second (third) term in line one describes the unwanted coupling
between cavity 2 and the $\left\vert e\right\rangle \rightarrow \left\vert
f\right\rangle $\ $\left( \left\vert g\right\rangle \rightarrow \left\vert
e\right\rangle \right) $\ transition with coupling constant $g_{2}^{\prime }$%
\ ($g_{2}^{\prime \prime }$) and detuning $\Delta _{2}^{\prime }=\omega
_{fe}-\omega _{c_{2}}$\ ($\Delta _{2}^{\prime \prime }=\omega _{eg}-\omega
_{c_{2}}$), the second (third) term in line two describes the unwanted
coupling between cavity $1^{\prime }$\ and the $\left\vert g\right\rangle
\rightarrow \left\vert f\right\rangle $\ $\left( \left\vert g\right\rangle
\rightarrow \left\vert e\right\rangle \right) $\ transition with coupling
constant $\mu _{1^{\prime }}^{\prime }$\ ($\mu _{1^{\prime }}^{\prime \prime
}$) and detuning $\Delta _{1^{\prime }}^{\prime }=\omega _{fg}-\omega
_{c_{1^{\prime }}}$\ ($\Delta _{1^{\prime }}^{\prime \prime }=\omega
_{eg}-\omega _{c_{1^{\prime }}}$), and the second (third) term in line three
describes the unwanted coupling between cavity $2^{\prime }$\ and the $%
\left\vert g\right\rangle \rightarrow \left\vert f\right\rangle $\ $\left(
\left\vert g\right\rangle \rightarrow \left\vert e\right\rangle \right) $\
transition with coupling constant $\mu _{2^{\prime }}^{\prime }$\ ($\mu
_{2^{\prime }}^{\prime \prime }$) and detuning $\Delta _{2^{\prime
}}^{\prime }=\omega _{fg}-\omega _{c_{2^{\prime }}}$\ ( $\Delta _{2^{\prime
}}^{\prime \prime }=\omega _{eg}-\omega _{c_{2^{\prime }}}$). In addition, $%
\varepsilon $\ is the Hamiltonian describing the inter-cavity crosstalk,
which is given by
\begin{eqnarray}
\varepsilon &=&\left( g_{12}e^{i\Delta _{12}t}\hat{a}_{1}^{+}\hat{a}_{2}+%
\text{h.c.}\right) +\left( g_{1^{\prime }2^{\prime }}e^{i\Delta _{1^{\prime
}2^{\prime }}t}\hat{b}_{1^{\prime }}^{+}\hat{b}_{2^{\prime }}+\text{h.c.}%
\right)  \notag \\
&&+\left( g_{11^{\prime }}e^{i\Delta _{11^{\prime }}t}\hat{a}_{1}^{+}\hat{b}%
_{1^{\prime }}+\text{h.c.}\right) +\left( g_{12^{\prime }}e^{i\Delta
_{12^{\prime }}t}\hat{a}_{1}^{+}\hat{b}_{2^{\prime }}+\text{h.c.}\right)
\notag \\
&&+\left( g_{21^{\prime }}e^{i\Delta _{21^{\prime }}t}\hat{a}_{2}^{+}\hat{b}%
_{1^{\prime }}+\text{h.c.}\right) +\left( g_{22^{\prime }}e^{i\Delta
_{22^{\prime }}t}\hat{a}_{2}^{+}\hat{b}_{2^{\prime }}+\text{h.c.}\right) ,
\end{eqnarray}%
where $g_{kl}$\ is the crosstalk strength between the two cavities $k$\ and $%
l$\ while $\Delta _{kl}=\omega _{c_{k}}-\omega _{c_{l}}$\ is the frequency
difference of the two cavities $k$\ and $l$\ $\left( k,l\in \{1,2,1^{\prime
},2^{\prime }\};k\neq l\right) $.

On the other hand, when the unwanted couplings between the qutrit's level
transitions and cavity $1^{\prime }$\ are considered, the Hamiltonian $H_{3}$%
\ in Eq. (9) is modified as%
\begin{eqnarray}
H_{3}^{\prime } &=&\left( \widetilde{g}e^{i\widetilde{\Delta }t}\hat{a}%
_{1^{\prime }}\sigma _{eg}^{+}+\widetilde{g}^{\prime }e^{i\widetilde{\Delta }%
^{\prime }t}\hat{a}_{1^{\prime }}\sigma _{fg}^{+}+\widetilde{g}^{\prime
\prime }e^{i\widetilde{\Delta }^{\prime \prime }t}\hat{a}_{1^{\prime
}}\sigma _{fe}^{+}+\text{h.c.}\right)  \notag \\
&&+\left( \Omega _{p}e^{-i\left[ \left( \omega _{p}-\omega _{eg}\right)
t+\phi \right] }\sigma _{eg}^{+}+\Omega _{p}^{\prime }e^{-i\left[ \left(
\omega _{p}-\omega _{fg}\right) t+\phi \right] }\sigma _{fg}^{+}+\Omega
_{p}^{\prime \prime }e^{-i\left[ \left( \omega _{p}-\omega _{fe}\right)
t+\phi \right] }\sigma _{fe}^{+}+\text{h.c.}\right) \text{,}  \notag \\
&&
\end{eqnarray}%
where the second (third) term in line one describes the unwanted coupling
between cavity $1^{\prime }$\ and the $\left\vert g\right\rangle \rightarrow
\left\vert f\right\rangle $\ $\left( \left\vert e\right\rangle \rightarrow
\left\vert f\right\rangle \right) $\ transition with coupling constant $%
\tilde{g}^{\prime }$\ ($\tilde{g}^{\prime \prime }$) and detuning $%
\widetilde{\Delta }^{\prime }=\omega _{fg}-\widetilde{\omega }_{c_{1^{\prime
}}}$\ ($\widetilde{\Delta }^{\prime \prime }=\omega _{fe}-\widetilde{\omega }%
_{c_{1^{\prime }}}$), while the second (third) term in line two describes
the unwanted coupling between the microwave pulse and the $\left\vert
g\right\rangle \rightarrow \left\vert f\right\rangle $\ $\left( \left\vert
e\right\rangle \rightarrow \left\vert f\right\rangle \right) $\ transition
with Rabi frequency $\Omega _{p}^{\prime }$ ($\Omega _{p}^{\prime \prime }$%
). For simplicity, we here assume that during the operation of step (v), all
other cavities are decoupled from the qutrit and the inter-cavity crosstalk
is negligible, which can be achieved by adjusting the cavity frequencies
such that the cavity-cavity frequency differences are sufficiently large.

During the operation of steps (iii) and (v), the dynamics of the lossy
system is determined by%
\begin{eqnarray}
\frac{d\rho }{dt} &=&-i\left[ H_{\mathrm{i}}^{\prime },\rho \right]
+\sum_{j=1}^{2}\kappa _{j}\mathcal{L}\left[ \hat{a}_{j}\right]
+\sum_{j^{\prime }=1^{\prime }}^{2^{\prime }}\kappa _{j^{\prime }}\mathcal{L}%
\left[ \hat{b}_{j^{\prime }}\right]  \notag \\
&&+\gamma _{eg}\mathcal{L}\left[ \sigma _{eg}^{-}\right] +\gamma _{fe}%
\mathcal{L}\left[ \sigma _{fe}^{-}\right] +\gamma _{fg}\mathcal{L}\left[
\sigma _{fg}^{-}\right]  \notag \\
&&+\gamma _{e,\varphi }\left( \sigma _{e}\rho \sigma _{e}-\sigma _{e}\rho
/2-\rho \sigma _{e}/2\right)  \notag \\
&&+\gamma _{f,\varphi }\left( \sigma _{f}\rho \sigma _{f}-\sigma _{f}\rho
/2-\rho \sigma _{f}/2\right) ,
\end{eqnarray}%
where $H_{\mathrm{i}}^{\prime }$\ is the modified Hamiltonian $H_{2}^{\prime
}$\ or $H_{3}^{\prime }$\ given above, $L\left[ \Lambda \right] =\Lambda
\rho \Lambda ^{+}-\Lambda ^{+}\Lambda \rho /2-\rho \Lambda ^{+}\Lambda /2$\
(with $\Lambda =\hat{a}_{j},,\hat{b}_{j^{\prime }},\sigma _{eg}^{-},\sigma
_{fe}^{-},\sigma _{fg}^{-})$,\ $\sigma _{eg}^{-}=\left\vert g\right\rangle
\left\langle e\right\vert ,$\ $\sigma _{fe}^{-}=\left\vert e\right\rangle
\left\langle f\right\vert ,$\ $\sigma _{fg}^{-}=\left\vert g\right\rangle
\left\langle f\right\vert ,$\ $\sigma _{e}=\left\vert e\right\rangle
\left\langle e\right\vert $, and $\sigma _{f}=\left\vert f\right\rangle
\left\langle f\right\vert .$\ In addition, $\kappa _{j}$\ ($\kappa
_{j^{\prime }}$) is the decay rate of cavity $j$\ ($j^{\prime }$);\ $\gamma
_{eg}$\ is the energy relaxation rate for the level $\left\vert
e\right\rangle $\ of the qutrit corresponding to the decay path $\left\vert
e\right\rangle \rightarrow \left\vert g\right\rangle $; $\gamma _{fe}$\ ($%
\gamma _{fg}$) is the relaxation rate for the level $\left\vert
f\right\rangle $\ of the qutrit associated with the decay path $\left\vert
f\right\rangle \rightarrow \left\vert e\right\rangle $\ ($\left\vert
f\right\rangle \rightarrow \left\vert g\right\rangle $); $\gamma _{e,\varphi
}$\ ($\gamma _{f,\varphi }$) is the dephasing rate of the level $\left\vert
e\right\rangle $\ ($\left\vert f\right\rangle $) of the qutrit.

For two SPS qubits, the Bell state is given by Eq. (16) with $c=d=1/\sqrt{2}%
, $ $n=2$\ and $n^{\prime }=2^{\prime },$\ i.e., the Bell state is $\frac{1}{%
\sqrt{2}}\left( \left\vert 0\right\rangle _{1}\left\vert 0\right\rangle
_{2}+\left\vert 1\right\rangle _{1}\left\vert 1\right\rangle _{2}\right) ,$
where the subscripts 1 and 2 represent SPS qubits 1 and 2, respectively. The
initial state of the whole system is $\frac{1}{\sqrt{2}}\left( \left\vert
0\right\rangle _{1}\left\vert 0\right\rangle _{2}+\left\vert 1\right\rangle
_{1}\left\vert 1\right\rangle _{2}\right) \left\vert \alpha \right\rangle
_{1^{\prime }}\left\vert \alpha \right\rangle _{2^{\prime }}\left\vert
g\right\rangle ,$ where the subscripts $1^{\prime }$\ and $2^{\prime }$\
represent CS qubits $1^{\prime }$\ and $2^{\prime }$, respectively.
According to Eq. (26), the ideal output state of the whole system after the
entire operation is
\begin{equation}
\left\vert g\right\rangle \left\vert 0\right\rangle _{1}\left\vert
-\right\rangle _{2}\otimes \frac{1}{\sqrt{2}}\left( \left\vert \alpha
\right\rangle _{1^{\prime }}\left\vert \alpha \right\rangle _{2^{\prime
}}+\left\vert -\alpha \right\rangle _{1^{\prime }}\left\vert -\alpha
\right\rangle _{2^{\prime }}\right) .
\end{equation}

The fidelity of the entire operation is defined as $F=\sqrt{\left\langle
\psi _{id}\right\vert \rho \left\vert \psi _{id}\right\rangle },$ where $%
\left\vert \psi _{id}\right\rangle $\ is the ideal output state given by Eq.
(34), while $\rho $\ is the final density matrix obtained by numerically
solving the master equation in a realistic situation.

For a flux qutrit, the typical transition frequency between neighboring
levels can be made as 1 to 20 GHz. As an example, consider $\omega
_{eg}/2\pi =8.0$\ GHz, $\omega _{fe}/2\pi =12.0$\ GHz, and $\omega
_{fg}/2\pi =20.0$\ GHz. Choose $\omega _{c_{1}}/2\pi =5.0$\ GHz for cavity $1
$. With a choice of $\Delta _{2}/2\pi =-250$\ MHz, $\Delta _{1^{\prime
}}/2\pi =-125$\ MHz, and $\Delta _{2^{\prime }}/2\pi =125$\ MHz, the cavity
frequencies are $\omega _{c_{2}}/2\pi =20.25$\ GHz, $\omega _{c_{1^{\prime
}}}/2\pi =12.125$\ GHz, and $\omega _{c_{2^{\prime }}}/2\pi =11.875$\ GHz.
Thus, we have $\Delta _{2}^{\prime }/2\pi =-8.25$\ GHz, $\Delta _{2}^{\prime
\prime }/2\pi =-12.25$\ GHz, $\Delta _{1^{\prime }}^{\prime }/2\pi =7.875$\
GHz, $\Delta _{1^{\prime }}^{\prime \prime }/2\pi =-4.125$\ GHz, $\Delta
_{2^{\prime }}^{\prime }/2\pi =8.125$\ GHz, and $\Delta _{2^{\prime
}}^{\prime \prime }/2\pi =-3.875$\ GHz. For the values of the cavity
frequencies here, a simple calculation gives $\Delta _{12}/2\pi =-15.25$\
GHz, $\Delta _{1^{\prime }2^{\prime }}/2\pi =0.25$\ GHz, $\Delta
_{11^{\prime }}/2\pi =-7.125$\ GHz, $\Delta _{12^{\prime }}/2\pi =-6.875$\
GHz, $\Delta _{21^{\prime }}/2\pi =8.125$\ GHz, and $\Delta _{22^{\prime
}}/2\pi =8.375$\ GHz. In addition, we have $\mu _{1^{\prime }}=$\ $\mu
_{2^{\prime }}=g_{2}/\sqrt{2}$\ according to Eq. (28).

We choose $\widetilde{\Delta }/2\pi =160$\ MHz, resulting in $\widetilde{%
\omega }_{c_{1^{\prime }}}/2\pi =7.84$\ GHz. For the value of $\widetilde{%
\omega }_{c_{1^{\prime }}}$\ here and the values of $\omega _{eg}$, $\omega
_{fe}$, $\omega _{fg}$\ given above, we have $\widetilde{\Delta }^{\prime
}/2\pi =12.16$\ GHz and $\widetilde{\Delta }^{\prime \prime }/2\pi =4.16$\
GHz. We choose $\widetilde{g}/2\pi =24.1$\ MHz, and thus $\Omega _{p}/2\pi
\sim 0.454$\ MHz (calculated according to Eq. (29) for $m=1$ and $\alpha
=1.86$). Because of $\omega _{p}=\omega _{eg}+\widetilde{g}^{2}/\widetilde{%
\Delta },$\ the driving field frequency $\omega _{p}$\ can be calculated
given the values of $\omega _{eg},$\ $\tilde{g},$\ and $\tilde{\Delta}$\
here.

The dipole matrix element $\chi _{fg}$\ between the two levels $\left\vert
g\right\rangle $\ and $\left\vert f\right\rangle ,$\ the dipole matrix
element $\chi _{fe}$\ between the two levels $\left\vert e\right\rangle $\
and $\left\vert f\right\rangle ,$\ and the dipole matrix element $\chi _{eg}$%
\ between the two levels $\left\vert g\right\rangle $\ and $\left\vert
e\right\rangle $\ are assumed to have the relationship of $\chi _{fg}\sim
\chi _{fe}\sim 4\chi _{eg}$, which can be made with appropriate design of
the qutrit system [43]. Thus, we choose $g_{2}^{\prime }\sim g_{2},$\ $%
g_{2}^{\prime \prime }\sim 0.25g_{2};$\ $\mu _{1^{\prime }}^{\prime }\sim
\mu _{1^{\prime }},$\ $\mu _{1^{\prime }}^{\prime \prime }\sim 0.25\mu
_{1^{\prime }};$\ $\mu _{2^{\prime }}^{\prime }\sim \mu _{2^{\prime }},$\ $%
\mu _{2^{\prime }}^{\prime \prime }\sim 0.25\mu _{2^{\prime }}$; $\widetilde{%
g}^{\prime }\sim 4\widetilde{g},$\ $\widetilde{g}^{^{\prime \prime }}\sim 4%
\widetilde{g};$\ and $\Omega _{p}^{\prime }\sim 4\Omega _{p},$\ $\Omega
_{p}^{^{\prime \prime }}\sim 4\Omega _{p}.$

Other parameters used in the numerical simulation are: (i) $\gamma
_{eg}^{-1}=4T$\ $\mu $s, $\gamma _{fe}^{-1}=2T$\ $\mu $s, $\gamma
_{fg}^{-1}=T$\ $\mu $s, (ii) $\gamma _{\phi e}^{-1}=\gamma _{\phi f}^{-1}=T$%
\ $\mu $s, (iii) $\kappa _{1},\kappa _{2},\kappa _{1^{\prime }},\kappa
_{2^{\prime }}=\kappa ,$ and (iv) $\alpha =1.86$. In addition, choose $%
g_{kl}=0.01g_{m}$\ [14], with $g_{m}=\max \{g_{2},\mu _{1^{\prime }},\mu
_{2^{\prime }}\}.$

\begin{figure}[tbp]
\begin{center}
\includegraphics[width=10.5 cm, clip]{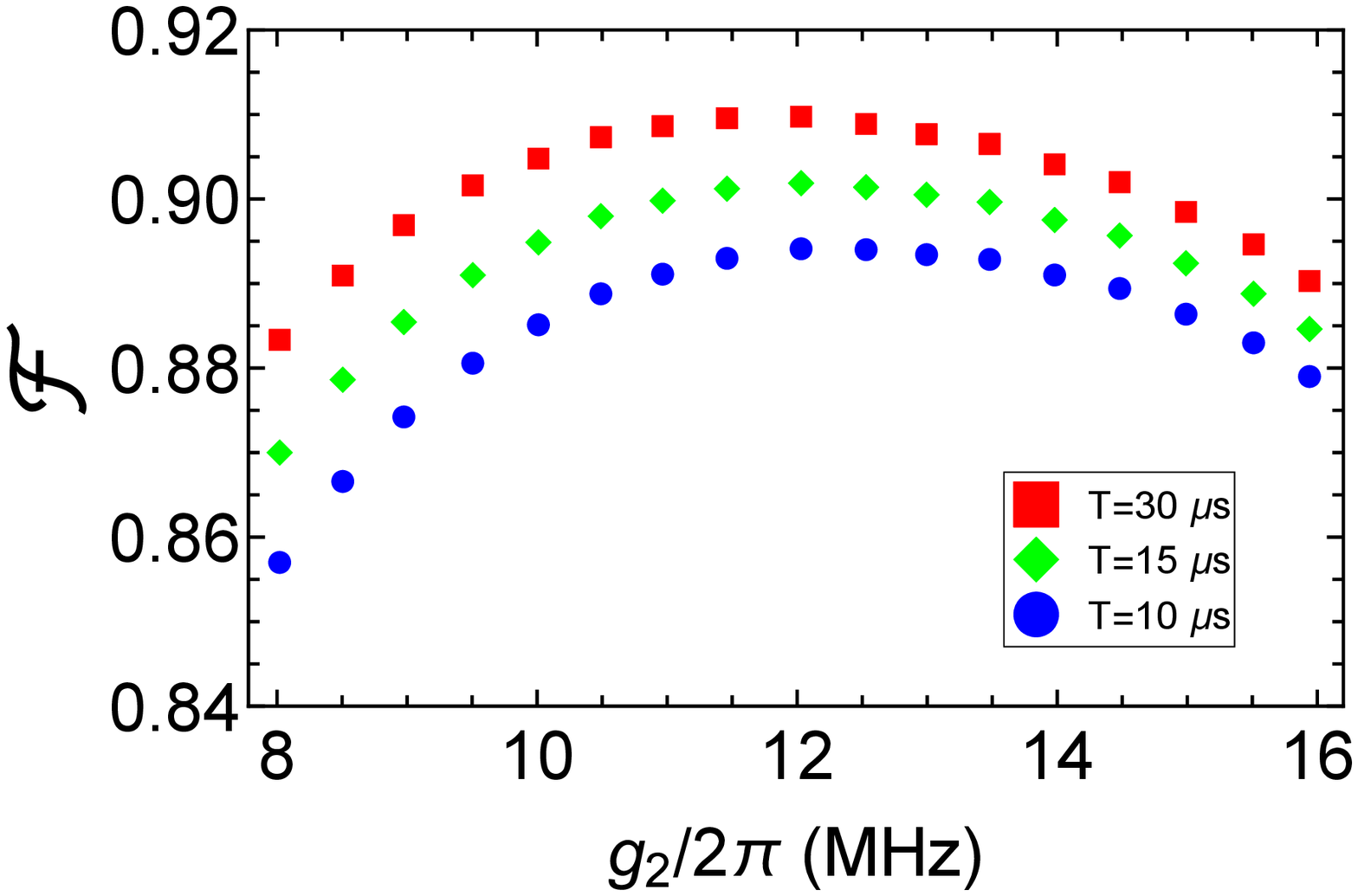} \vspace*{-0.08in}
\end{center}
\caption{(color online) Fidelity versus $g_{2}$. The parameters used in the
numerical simulation are referred to the text.}
\label{fig:3}
\end{figure}

\begin{figure}[tbp]
\begin{center}
\includegraphics[width=10.5 cm, clip]{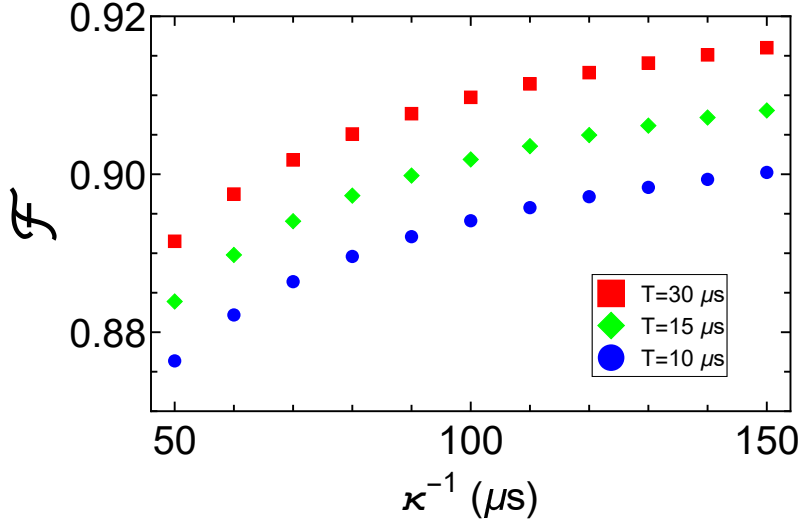} \vspace*{-0.08in}
\end{center}
\caption{(color online) Fidelity versus $\protect\kappa ^{-1}$. The
parameters used in the numerical simulation are referred to the text.}
\label{fig:4}
\end{figure}

By numerically solving the master equation (33), Fig. 3 is plotted, which
illustrates the fidelity versus $g_{2}$\ for $\kappa ^{-1}=100$\ $\mu $s and
$T=10$ $\mu $s$,$\ $15$ $\mu $s$,$\ $30$\ $\mu $s. Note that all other
coupling constants can be determined for a given $g_{2}$, as described
above. In Fig. 3, the optimal values of $g_{2}$, corresponding to the peaks
of three curve lines, are $2\pi \times 12.03$\ MHz for $T=10$ $\mu $s, $15$ $%
\mu $s and $30$ $\mu $s. Next, we plot Fig. 4, which shows the fidelity
versus $\kappa ^{-1}$\ for $g_{2}=2\pi \times 12.03$\ MHz and $T=10$ $\mu $s$%
,$\ $15$ $\mu $s$,$\ $30$\ $\mu $s. From Fig. 4, one can see that the
fidelity increases with $\kappa ^{-1}$\ and $T$, and for $T\geq 15$\ $\mu $s
and $\kappa ^{-1}\geq 100$\ $\mu $s, the fidelity exceeds $90.2\%$, impling
the high-fidelity Bell-state transfer from two SPS qubits onto two CS
qubits. It should be mentioned that when taking into account the qutrit
decoherence, the cavity decay, and the inter-cavity crosstalk in the
operations of steps (i), (ii), (iv) and (vi), the fidelity would be slightly
decreased but will not be greatly decreased since these steps of operation
can be completed very fast due to the use of the resonant interactions.

The coupling constants are readily achievable because a coupling constant $%
\sim 2\pi \times 636$\ MHz has been reported for a flux device coupled to a
microwave cavity [44]. For the parameters chosen above, the state transfer
time is estimated as $\sim 1.41$ $\mu $s, which is much shorter than the
decoherence times of the coupler qutrit ($10$\ $\mu $s $-$\ $120$\ $\mu $s)
and the cavity decay times ($50$\ $\mu $s $-$\ $150$\ $\mu $s) considered in
Figs. 3 and 4. For decoherence time of the flux qutrit, a rather
conservative case is considered in the numerical simulation, because
decoherence time 70 $\mu $s to 1 ms for a superconducting flux device has
been reported in experiments [45,46]. For $\kappa ^{-1}=100$\ $\mu $s and
the cavity frequencies $\omega _{c_{1}},\omega _{c_{2}},\omega
_{c_{1^{\prime }}},\omega _{c_{2^{\prime }}}$\ given above, a simple
calculation gives $Q_{1}\sim 3.14\times 10^{6}$\ for cavity $1,$\ $Q_{2}\sim
1.27\times 10^{7}$\ for cavity $2,$\ $Q_{1^{\prime }}\sim 7.61\times 10^{6}$%
\ for cavity $1^{\prime },$\ and $Q_{2^{\prime }}\sim 7.46\times 10^{6}$\
for cavity $2^{\prime }.$\ Note that a high quality factor $Q=3.5\times
10^{7}$\ of a 3D microwave cavity has been experimentally demonstrated [47].
Our analysis here shows that the high-fidelity transfer of the Bell state of
two SPS qubits onto two CS qubits is feasible within the current circuit QED
technology.

\section{Conclusion}

We have explicitally shown how to transfer partially- or maximally-entangled
states (including Bell states and GHZ states) of $n$\ single-photon-state
(SPS)\ qubits onto $n$\ coherent-state (CS)\ qubits in a circuit-QED system.
As shown above, this proposal has these distinguishing features: (i) Only a
coupler qutrit is needed, hence the circuit hardware resources is
significantly reduced; (ii) The operation time does not increase with the
increasing of the number of qubits; (iii) Decoherence from the coupler
qutrit is greatly suppressed, because the higher energy level of the coupler
qutrit is not excited during the entire operation; and (iv) When the
dissipation is negligible, the state transfer is deterministic because of no
measurement being performed. Furthermore, we have numerically investigated
the experimental feasibility for transferring the Bell state of two SPS
qubits onto two CS qubits. The numerical simulation shows that high-fidelity
transfer of the Bell state, from two SPS qubits onto two CS qubits, can be
achieved with current circuit-QED technology.

Finally, it should be mentioned that since all basic operations are unitary
and measurement is not involved, the entangled states of SPS qubits can be
transferred back onto CS qubits by performing reverse operations when the
dissipation is negligible. This proposal can be applied to transfer
maximally- or partially-entangled states between $n$ SPS qubits and $n$ CS
qubits in a wide range of physical systems, such as $2n$\ microwave or
optical cavities coupled to a natural or artificial atom.

\section*{Acknowledgements}

This work was partly supported by the National Natural Science Foundation of
China (NSFC) (11074062, 11374083, 11774076), the Key-Area Research and
Development Program of GuangDong province (2018B030326001), and the NKRDP of
China (2016YFA0301802).

\end{document}